\documentclass[twocolumn,prl,showpacs,
               superscriptaddress,amsmath,amssymb]{revtex4}

\usepackage{graphicx}   
\usepackage{dcolumn}    
\usepackage{bm}         
\usepackage{longtable}  

\newcommand{\ie}{{\em i.e.\/}}

\newcommand{\dfra}{\displaystyle\frac}

\def \ie {{\it i.e.}}

\begin{document}


\title{Retrieving the missed particle-antiparticle degrees of freedom
       of Dirac particles}

\author{Shun-Jin Wang}
 \affiliation{Department of Physics, Sichuan University,
              Chengdu 610064, China}
 \affiliation{Center of Theoretical Nuclear Physics, National Laboratory
              of Heavy Ion Accelerator, Lanzhou 730000, China}
\author{Shan-Gui Zhou}
 \email{sgzhou@itp.ac.cn}
 \homepage{http://www.itp.ac.cn/~sgzhou}
 \affiliation{Institute of Theoretical Physics, Chinese Academy Sciences,
              Beijing 100080, China}
 \affiliation{Center of Theoretical Nuclear Physics, National Laboratory
              of Heavy Ion Accelerator, Lanzhou 730000, China}
\author{Hans-Christian Pauli}
 \affiliation{Max-Planck-Institut f\"ur Kernphysik,
              69029 Heidelberg, Germany}

\date{\today}

\begin{abstract}
The missed particle-antiparticle degrees of freedom are retrieved
and the corresponding particle-antiparticle intrinsic space are
introduced to study the dynamical symmetry of the Dirac particle.
As a result, the particle-antiparticle quantum number appears
naturally and the Dirac particle has five quantum numbers instead
of four. An anti-symmetry (different from the conventional
symmetry) of the Dirac Hamiltonian and a dual symmetry of its
eigen functions are explored. The $\hat{\kappa}$ operator of the
Dirac equation in central potentials is found to be the analog of
the helicity operator of the free particle---the alignment of the
spin along the angular momentum.
\end{abstract}


\maketitle


\textit{Introduction.}The eigen solutions of the Dirac equation
for a free particle and/or a particle in central potentials are
the most basic elements of the relativistic quantum
mechanics~\cite{bd64} and quantum field theory~\cite{iz80}.
Although the particle and antiparticle states of a Dirac particle
are conventionally considered as two states of the same Fermion,
and the anti-commutators and the charge conjugate transformation
(the $\mathcal{C}$ operator) for the field operators are
introduced, the particle-antiparticle quantum number is still
missing and the $\mathcal{C}$ transformation just belongs to a
discrete two-element group. This unpleasant situation occurs due
to the veiled reason that the particle-antiparticle degrees of
freedom are hidden in the $\gamma$ matrices of the Dirac equation.
In this Letter we shall show that the $\gamma$ matrices can be
split into products of two parts which correspond to the spin
operators and the particle-antiparticle operators respectively.
The dynamical degrees of freedom (DoF) for a Dirac particle thus
become complete and include those in coordinate space, spin space,
and particle-antiparticle space. As a result, the Dirac particle
has five quantum numbers instead of four. The splitting of
$\gamma$ matrices and the introduction of particle-antiparticle
operators are necessary for a correct description of interactions
and may have profound meaning.


\textit{The dynamical degrees of freedom for free Dirac
particles.} The Hamiltonian of a free Dirac particle reads
\begin{equation}
 \widehat{H} = c \bm{\alpha } \cdot \bm{p} + mc^{2}\beta.
 \label{eq:DHFree0}
\end{equation}
Introducing the particle-antiparticle operators $\bm\tau$ as
follows
\begin{equation}
 \tau_{1} = \left(
             \begin{array}{cc}
              0 & 1 \\
              1 & 0
             \end{array}
            \right) ,\
 \tau_{2} = \left(
             \begin{array}{cc}
              0 & -i \\
              i & 0
             \end{array}
            \right) ,\
 \tau_{3} = \left(
             \begin{array}{cc}
              1 & 0 \\
              0 & -1
             \end{array}
            \right) ,
\end{equation}
the matrices $\bm\alpha$ and $\beta$ (or $\gamma_{\mu}$) can be
written as direct products of the spin operator $\bm\sigma$ and
the particle-antiparticle operator $\bm\tau$
\begin{equation}
 \bm{\alpha} = \bm{\sigma}  \otimes \tau_{1} ,\
       \beta = \mathbb{I}_{\sigma} \otimes \tau_{3} .
\end{equation}
Here the convention is used that the spin operators
$(\mathbb{I}_{\sigma},\sigma_{i})$ should be inserted into the
particle-antiparticle operators $(\mathbb{I}_{\tau},\tau_{i})$ as
their operator-valued elements. It is found that the algebra
$(\mathbb{I}_{\sigma},\sigma_{i})\otimes(\mathbb{I}_{\tau},\tau_{i})$
is equivalent to the Clifford algebra or Dirac ring which is
indispensible for a proper description of
interations~\cite{ro60,Lee83} and also justifies the above
splitting scheme. Hence the Hamiltonian can be rewritten as
\begin{equation}
 \widehat{H} = c \bm{p} \cdot \bm{\sigma} \otimes \tau_{1}
             + m c^{2} \mathbb{I}_{\sigma} \otimes \tau_{3}.
 \label{eq:DHFree}
\end{equation}

Now it is evident that a free Dirac particle has three types of
DoF's: the DoF's in coordinate space $\bm{r}$, spin space $s$, and
particle-antiparticle space $\tau$. Equation~(\ref{eq:DHFree})
clearly shows that the Dirac Hamiltonian possesses the dynamical
group $\text{SU}_{\sigma}(2) \otimes \text{SU}_{\tau }(2)$. Thus a
free Dirac particle should have five mutually commutative
invariant operators and five quantum numbers (instead of four)
which uniquely specify its quantum states. The five operators are:
(i) three components of the conserved momentum $\bm{p}$ ($[\bm{p},
\widehat{H}]=0$); (ii) since the momentum is conserved, the
helicity can be defined as $\sigma_{\parallel} = (\bm{p} \cdot
\bm{\sigma})/p$ which commutes with the Hamiltonian; (iii)
finally, the particle-antiparticle operator reads
\begin{equation}
 \widetilde{\tau}_{3}
 = \dfra{1}{E} (\mu\ cp \tau_1 + mc^2 \tau_3 )
 = \dfra{1}{E}
   \left(
    \begin{array}{cc}
     mc^{2}  & \mu\ cp \\
     \mu\ cp & -mc^{2}
    \end{array}
   \right),
\end{equation}
where $\mu$ is the eigen value of $\sigma_\parallel$, and the
commutation relation $[\widetilde{\tau}_{3},\widehat{H}]=0$ holds
too.

The eigen wave function for a free Dirac particle is a product of
three separable parts
\begin{equation}
 \textstyle
 \widetilde{\Psi}_{p \mu \nu }(\bm{r},s,\tau )
 = \frac{1}{(2\pi \hbar )^{3/2}}
   e^{i\bm{p} \cdot \bm{r}}
   \widetilde{U}(\mu)
   \widetilde{V}(\nu)
 \label{eq:WFFree}
\end{equation}
where the spin wave function $\widetilde{U}(\mu)$ is the eigen
function of $\sigma_{\parallel}$
\begin{equation}
 \sigma_{\parallel} \widetilde{U}(\mu )
 = \mu \widetilde{U}(\mu ),
 \text{ with }\mu =\pm 1.
\end{equation}
If $\sigma_{\parallel}$ is diagonal, \ie,
 $\sigma_{\parallel} = \left(
                       \begin{array}{cc}
                        1 & 0 \\
                        0 & -1
                       \end{array}
                      \right)$,
then we have
\begin{equation}
 \widetilde{U}(+1) = \left(
                      \begin{array}{c}
                       1 \\
                       0
                      \end{array}
                     \right) ,\
 \widetilde{U}(-1) = \left(
                      \begin{array}{c}
                       0 \\
                       1
                      \end{array}
                     \right) .
 \label{eq:WFFree1}
\end{equation}

In the above helicity representation, the eigen functions of
$\widetilde{\tau}_{3}$ with the eigen values $+1$ and $-1$
corresponding to particle and antiparticle respectively, read
\begin{equation}
 \textstyle
 \widetilde{V}(+1) = \sqrt{\frac{E+mc^{2}}{2E}}
                     \left(
                      \begin{array}{c}
                       1 \\
                       \frac{\mu cp}{E+mc^{2}}
                      \end{array}
                     \right) ,
 \label{eq:WFFree2}
\end{equation}
\begin{equation}
 \textstyle
 \widetilde{V}(-1) = i\sqrt{\frac{E+mc^{2}}{2E}}
                      \left(
                       \begin{array}{c}
                        \frac{-\mu cp}{E+mc^{2}} \\
                        1
                       \end{array}
                      \right) ,
 \label{eq:WFFree3}
\end{equation}
where $E=\sqrt{m^{2}c^{4}+c^{2}p^{2}}$. Inserting
Eqs.~(\ref{eq:WFFree1}, \ref{eq:WFFree2}, \ref{eq:WFFree3}) into
Eq.~(\ref{eq:WFFree}), we recover the standard solutions for a
free Dirac particle.

In the above representation, the $4\times 4$ helicity operator is
also diagonal,
\begin{equation}
 \widetilde{\Sigma}_{\parallel}
 = \mathbb{I}_{\tau}\otimes \sigma_z ,\ \
 \widetilde{\Sigma}_{\parallel}
 \widetilde{\Psi }_{p\mu \nu}(\bm{r},s,\tau)
 = \mu \widetilde{\Psi }_{p\mu \nu}(\bm{r},s,\tau).
\end{equation}
and the particle-antiparticle operator $\widetilde{\tau}_{3}$
becomes a $4\times 4$ matrix
\begin{eqnarray}
 \widetilde{T}_{3}
 & = &
  \dfra{1}{E}
  \left(
   \begin{array}{cc}
    mc^{2}         &  cp \sigma _{z} \\
    cp \sigma _{z} & -mc^{2}
   \end{array}
  \right) ,
\end{eqnarray}
with
\begin{equation}
 \widetilde{T}_{3}
 \widetilde{\Psi}_{p\mu \nu}(\bm{r},s,\tau)
 = \nu \widetilde{\Psi}_{p\mu \nu}(\bm{r},s,\tau).
\end{equation}

It turns out that the Hamiltonian is proportional to the
particle-antiparticle operator,
\begin{equation}
 \widetilde{H} = E \widetilde{T}_{3},\ \
 \widetilde{H} \widetilde{\Psi}_{p\mu \nu}(\bm{r},s,\tau)
 = \nu E \widetilde{\Psi}_{p\mu \nu}(\bm{r},s,\tau) ,
\end{equation}
which tells us that the negative energies of the antiparticles are
in fact from the negative antiparticle quantum number $-1$.

We find that a particle state is transformed to an antiparticle
state with the helicity intact by the following unitary
transformation,
\begin{equation}
 \widetilde{C}
 = \left(
    \begin{array}{cc}
     0              & i \mathbb{I} \\
    -i \mathbb{I}   & 0
    \end{array}
   \right)
 = \widetilde{C}^{\dagger} = \widetilde{C}^{-1}
\end{equation}
\begin{equation}
 \widetilde{C}\widetilde{\Psi}_{p\mu \nu}(\bm{r},s,\tau)
 = \widetilde{\Psi}_{p\mu -\nu}(\bm{r},s,\tau) ,
\end{equation}
and {\it vice versa}. Thus we can call $\widetilde C$ the
particle-antiparticle transformation. The Hamiltonian
$\widetilde{H}$ is {\it anti-symmetric} under the
particle-antiparticle transformation,
\begin{equation}
 \widetilde{C} \widetilde{H} \widetilde{C}^{-1}
 = -\widetilde{H} ,\ \text{\ie, }\ \{\widetilde{H},\widetilde{C}\} = 0.
\end{equation}

By virtue of the unitary transformation,
\begin{equation}
 \textstyle
 \widehat{u}
 =\mathbb{I}_{\tau}\otimes
   \sqrt{\frac{1+\widehat{p}_z}{2}}
   \left(
    \begin{array}{cccc}
     1 & \frac{\widehat{p}_r}{1+\widehat{p}_z} \\
     \frac{\widehat{p}_l}{1+\widehat{p}_z} & -1 \\
     \end{array}
\right) ,
\end{equation}
where $\widehat{p}_{z}=p_{z}/p$,
$\widehat{p}_{r}=(p_{x}-ip_{y})/p$, and
$\widehat{p}_{l}=(p_{x}-ip_{y})/p$, we obtain the eigen solutions
of the Dirac equation in the Dirac-Pauli $\gamma$-matrix
representation as follows
\begin{equation}
 \Psi _{p\mu \nu}(\bm{r},s,\tau)
 = \widehat{u} \widetilde{\Psi}_{p\mu \nu}.
\end{equation}
The eigen solutions of the Dirac equation are
\begin{equation}
 \textstyle
 \Psi_{p++}(\bm{r},s,\tau )
 =
   \mathcal{N}
   \left(
    \begin{array}{c}
     1 \\
     \frac{\widehat{p}_l}{1+\widehat{p}_z} \\
     \frac{cp}{E+mc^{2}} \\
     \frac{\widehat{p}_l}{1+\widehat{p}_z} \frac{cp}{E+mc^{2}}
    \end{array}
   \right)
   e^{i\bm{p} \cdot \bm{r}},
\end{equation}
\begin{equation}
 \textstyle
 \Psi_{p-+}(\bm{r},s,\tau )
 =
   \mathcal{N}
   \left(
    \begin{array}{c}
     \frac{\widehat{p}_r}{1+\widehat{p}_z} \\
    -1 \\
     \frac{-\widehat{p}_r}{1+\widehat{p}_z} \frac{cp}{E+mc^{2}} \\
     \frac{cp}{E+mc^{2}}
    \end{array}
   \right)
   e^{i\bm{p} \cdot \bm{r}},
\end{equation}
\begin{equation}
 \textstyle
 \Psi_{p+-}(\bm{r},s,\tau )
 = i
   \mathcal{N}
   \left(
    \begin{array}{c}
     \frac{-cp}{E+mc^2} \\
     \frac{\widehat{p}_l}{1+\widehat{p}_z} \frac{cp}{E+mc^2} \\
     1 \\
     \frac{\widehat{p}_l}{1+\widehat{p}_z}
    \end{array}
   \right)
   e^{i\bm{p} \cdot \bm{r}},
\end{equation}
\begin{equation}
 \textstyle
 \Psi_{p--}(\bm{r},s,\tau )
 = i
   \mathcal{N}
   \left(
    \begin{array}{c}
     \frac{\widehat{p}_r}{1+\widehat{p}_z} \frac{cp}{E+mc^{2}} \\
     \frac{-cp}{E+mc^{2}} \\
     \frac{\widehat{p}_r}{1+\widehat{p}_z} \\
    -1
    \end{array}
   \right)
   e^{i\bm{p} \cdot \bm{r}} ,
\end{equation}
where
\begin{equation}
 \textstyle
 \mathcal{N} =
   \frac{1}{(2\pi \hbar )^{3/2}}
   \sqrt{\frac{E+mc^2}{2E}}
   \sqrt{\frac{1+\widehat{p}_z}{2}} .
\end{equation}

The $4\times 4$ helicity operator now becomes
\begin{equation}
 \Sigma_{\parallel} =\mathbb{I}_{\tau}\otimes
 \left(
  \begin{array}{cccc}
   \widehat{p}_{z} &  \widehat{p}_{r} \\
   \widehat{p}_{l} & -\widehat{p}_{z} \\
 \end{array}
 \right) ,
\end{equation}
with
\begin{equation}
 \Sigma _{\parallel}\Psi_{p\mu \nu}(\bm{r},s,\tau)
 = \mu \Psi_{p\mu \nu}(\bm{r},s,\tau).
\end{equation}

The $4\times 4$ particle-antiparticle operator now reads
\begin{equation}
 T_{3} =
 \dfra{1}{E}
 \left(
  \begin{array}{cccc}
   mc^2 & 0 & cp\ \widehat{p}_z &  cp\ \widehat{p}_r \\
   0 & mc^2 & cp\ \widehat{p}_l & -cp\ \widehat{p}_z \\
   cp\ \widehat{p}_z & cp\ \widehat{p}_r & -mc^2 & 0 \\
   cp\ \widehat{p}_l & -cp\ \widehat{p}_z & 0 & -mc^2
  \end{array}
 \right) ,
\end{equation}
with
\begin{equation}
 T_{3}\Psi_{p\mu\nu}(\bm{r},s,\tau )
 = \nu \Psi_{p\mu \nu}(\bm{r},s,\tau).
\end{equation}

The Hamiltonian of the Dirac equation is $ \widehat{H} = ET_{3}$,
and
\begin{equation}
 \widehat{H}\Psi_{p\mu \nu}(\bm{r},s,\tau)
 = \nu E\Psi_{p\mu \nu}(\bm{r},s,\tau) .
\end{equation}

The particle-antiparticle unitary transformation now reads
\begin{equation}
 C = \widehat{u} \widetilde{C} \widehat{u}^{-1}
   = -\mathbb{I}_{\sigma}\otimes \tau_2,
\end{equation}
which transforms a particle state to an antiparticle state with
the helicity intact, and {\it vice versa},
\begin{equation}
 C \Psi_{p\mu\nu}(\bm{r},s,\tau)
 = \Psi_{p\mu -\nu}(\bm{r},s,\tau).
\end{equation}

Again, the Hamiltonian $\widehat{H}$ is anti-commutative with the
particle-antiparticle transformation $C$, $C \widehat{H} C^{-1} =
-\widehat{H}$, and $\{\widehat{H},C\}=0$.



\textit{Dirac particles in central potentials: (A)
$\widehat{\kappa}$ --- the analog of the helicity in the central
potential.} The Hamiltonian of a Dirac particle in a central
vector potential $V_{v}(r)$ reads
\begin{equation}
 \widehat{H} = c \bm{\alpha} \cdot \bm{p}
             + mc^{2} \beta + V_{v}(r).
\end{equation}

For a free Dirac particle, the conserved linear momentum of the
particle specifies a special direction of the system. The spin
alinment of the Dirac particle along the linear momentum results
in the helicity which is conserved consistently with the linear
momentum. The nature of the spin alignment is closely related to
the linearization of the squared relativistic energy operator
\begin{equation}
 [ \widehat{H} - V_{v}(r) ]^{2}
 = [ mc^{2} ]^{2} + (cp)^{2} ,\
 (cp)^{2} = -(c\hbar)^{2} \nabla ^{2} .
 \label{eq:DHL}
\end{equation}

For the free particle, the central potential $V_v(r)$ in
Eq.~(\ref{eq:DHL}) vanishes and
\begin{equation}
 \widehat{H}^{2} = [mc^{2}]^{2} + (cp)^{2}.
\end{equation}
Since the linear momentum is conserved, the linearization should
be carried out in Cartesian coordinates, and one gets the Dirac
Hamiltonian Eq.~(\ref{eq:DHFree0}). This means that the
linearization of the squared free particle Hamiltonian operator in
Cartesian coordinates leads to the helicity operator and  the
unique spin alignment of the free particle along its linear
momentum direction naturally leads to the helicity conservation.

In a central potential, the particle has certain angular momentum,
and the natural frame is the spherical coordinates. In this case,
the squared kinetic energy operator reads
\begin{equation}
 \textstyle
 (cp)^2 = -(c\hbar)^2 \nabla^2
          = -(c\hbar)^2 \left[ \frac{1}{r^2} \frac{d}{dr}r^2 \frac{d}{dr}
                                -\frac{\widehat{L}^2}{r^2}
                        \right] .
\end{equation}
The problem is how to linearize $[ \widehat{H} - V_{v}(r) ]^{2}$
in the spherical coordinates. This has been done and the result is
well known as follows,
\begin{equation}
 \textstyle
 \widehat{H} - V_{v}(r)
 = \beta mc^{2}
 + \alpha_{r}\left( \widehat{p}_{r} + \frac{ic\hbar}{r}\beta \widehat{\kappa}
             \right) ,
\end{equation}
\begin{eqnarray}
 \widehat{H}
 & = &
 \textstyle
 \alpha_{r}\left( c \widehat{p}_{r} + \frac{ic\hbar}{r} \beta \widehat{\kappa}
           \right)
  + V_{v}(r) + \beta mc^{2}
 \nonumber \\
 & = &
 \left(
  \begin{array}{cc}
   mc^{2}+V_{v} & c \widehat{p}_{r} \sigma_{r}
                 -\frac{ic\hbar}{r}\sigma_{r} \widehat{\kappa} \\
   c \widehat{p}_{r} \sigma_{r}
  +\frac{ic\hbar}{r} \sigma_{r}\widehat{\kappa} & -mc^{2}+V_{v}
  \end{array}
 \right),
\end{eqnarray}
where
\begin{equation}
 \textstyle
 \alpha_{r} = \frac{\widehat{\alpha} \cdot \widehat{r}}{r}
            = \left(
               \begin{array}{cc}
                0          & \sigma_{r} \\
                \sigma_{r} & 0
               \end{array}
              \right) ,
    \sigma_{r} = \frac{\bm{\sigma} \cdot \bm{r}}{r}
            = \sqrt{\frac{3}{4\pi}}\sum_{\mu} Y_{1 \mu}\sigma_{\mu},
\end{equation}
with
\begin{equation}
 \textstyle
 \sigma_{0} = \sigma_{z},\
 \sigma_{\pm} = \frac{1}{2} (\sigma_{x} \pm i\sigma y) ,\
 \widehat{p}_{r} = -i\hbar (\frac{d}{dr}+\frac{1}{r}),
\end{equation}
and $\widehat{\kappa }$ appearing in $\widehat{H}$ is just
referred to its eigen values $\pm (j+\frac{1}{2})$ symbolically,
it is in fact a $4\times 4$ operator
\begin{equation}
 \hbar \widehat{\kappa}
 = \beta (\bm{\Sigma} \cdot \bm{L} + \hbar)
 = \left(
    \begin{array}{cc}
     \bm{\sigma} \cdot \bm{L} + \hbar & 0 \\
     0 & -(\bm{\sigma} \cdot \bm{L} + \hbar)
    \end{array}
\right).
\end{equation}

This means that $\widehat{\kappa}$ comes from the linearization of
the operator $[ \widehat{H} - V_{v}(r) ]^{2}$  in the spherical
coordinates and $\widehat{\kappa}$ is thus the analog of the
helicity for the Dirac particle in the central potentials. In
fact, in the central potentials, the only characteristic direction
of the system is related to its angular momentum, so that the spin
can only make its alignment along the angular momentum with two
possibilities for a given total angular momentum $j$:
$j=l+\frac{1}{2}$ for parallel alignment and $j=(l+1)-\frac{1}{2}$
for anti-parallel alignment. The mixing between the harmonic spin
functions $\Phi_{jm}^{A} \equiv \sum_{m_l,m_s} C^{jm}_{lm_l1/2m_s}
Y_{lm_l}(\Omega)\chi_{m_s}(s)$ and $\Phi_{jm}^{B} \equiv
\sum_{m_l,m_s} C^{jm}_{l+1m_l1/2m_s}
Y_{l+1m_l}(\Omega)\chi_{m_s}(s)$ is due to the scalar operator
$\alpha_{r}$ contains the $1$-rank spherical harmonic function
$Y_{1\mu}$ and the $1$-rank tensors $\sigma_{\mu}(\mu=0,\pm 1)$
which change angular momentum by $\Delta l = \pm 1$ and keep the
total angular momentum $jm$ conserved.

Since in the $\gamma$-splitting scheme, the parity operator
$i\gamma_{4}=I_{\sigma}\otimes \tau_{3}$ indicates that the
particle and antiparticle components have opposite intrinsic
parities which, together with the opposite parities of the
components $\Phi_{jm}^{A}$ and $\Phi_{jm}^{B}$, make the Dirac
particle in central potentials have a definite total parity.


\textit{(B) The particle-antiparticle symmetry of Dirac equation
in central potentials.} To study the particle-antiparticle
symmetry of Dirac equation in central potentials, we add a scalar
central potential $V_{s}(r)$ in the Dirac Hamiltonian,
\begin{eqnarray}
 \widehat{H}
 & = &
 \textstyle
 \alpha_{r} \left[ c \widehat{p}_{r}
                  + \frac{ic\hbar}{r} \beta \widehat{\kappa} \right]
 + V_{v}(r) + \beta [ mc^{2} + V_{s}(r)]
 \nonumber \\
 & = &
 \left(
  \begin{array}{cc}
   (mc^{2}+V_{s})+V_{v} &
   c\widehat{p}_{r}\sigma_{r}-\frac{ic\hbar}{r}\sigma_{r}\widehat{\kappa} \\
   c\widehat{p}_{r}\sigma_{r}+\frac{ic\hbar}{r}\sigma_{r}\widehat{\kappa} &
   -(mc^{2}+V_{s})+V_{v}
  \end{array}
 \right).
\end{eqnarray}

Let's introduce the particle-antiparticle unitary transformation
\begin{equation}
 C = \left(
      \begin{array}{cc}
       0 & i\sigma _{r} \\
       -i\sigma _{r} & 0
      \end{array}
     \right) = C^{\dagger}=C^{-1},
\end{equation}
which is just a generalization of the unitary
particle-antiparticle transformation for the free Dirac particle.
The transformed Hamiltonian is
\begin{equation}
 \widetilde{H}
  = C \widehat{H} C^{-1} =  -\widehat{H}.
\end{equation}
Here we have used the transformation property of the scalar
potential and the vector potential as usual
\begin{equation}
  C V_{s}C^{-1}=V_{s}, \ \ C V_{v}C^{-1}=-V_{v}
\end{equation}
The Hamiltonian $\widehat{H}$ of the system is again
anti-commutative with the unitary particle-antiparticle
transformation $C$, $\{\widehat{H},C\}=0$.

Under the transformation $C$, $\widetilde{\Psi}_{njm\kappa\nu} =
C\Psi_{njm\kappa\nu}$, the Dirac equations for the particle and
antiparticle read,
\begin{equation}
 \widehat{H} \Psi_{njm\kappa\nu} = E \Psi_{njm\kappa\nu} ,
\end{equation}
\begin{equation}
 \widehat{H} \widetilde{\Psi}_{njm\kappa\nu}
 = -E \widetilde{\Psi}_{njm\kappa\nu}
 .
\end{equation}
Explicitly,
\begin{equation}
 \Psi_{njm++}=
  \left(
    \begin{array}{c}
     \Phi_{jm}^{A}  f(r) \\
     \Phi_{jm}^{B} ig(r)
    \end{array}
   \right) ,
 \Psi_{njm+-}
 = i \left(
      \begin{array}{c}
       \Phi_{jm}^{A}  g(r) \\
       \Phi_{jm}^{B} if(r)
      \end{array}
     \right) ,
\end{equation}
\begin{equation}
 \Psi_{njm-+}=
  \left(
    \begin{array}{c}
     \Phi_{jm}^{B}  f(r) \\
     \Phi_{jm}^{A} ig(r)
    \end{array}
   \right) ,
 \Psi_{njm--}
 = i \left(
      \begin{array}{c}
       \Phi_{jm}^{B}  g(r) \\
       \Phi_{jm}^{A} if(r)
      \end{array}
     \right) .
\end{equation}

Since the particle-antiparticle transformation exchanges the
radial functions $f(r)$ and $g(r)$ between the two components of
the eigen functions of the Dirac equation and keeps the total
angular momentum (spherical spinors) of the eigen functions
$\Phi_{jm}^{A}$ and $\Phi_{jm}^{B}$ intact, the transformed
Hamiltonian leads to the same coupled eigen equations for the
radial functions $f(r)$ and $g(r)$ of the antiparticle as that for
the particle. Therefore the antiparticle has the same radial eigen
functions as the particle does, but these two eigen radial
functions exchange their position in their eigen solutions.

In the above, besides the quantum number $\kappa$ as the eigen
values of $\hat{\kappa}/(j+1/2)$ ($\kappa=\pm 1$ instead of $\pm
(j+1/2)$), we have introduced the particle-antiparticle quantum
numbers $\nu$ ($=\pm 1$). The eigen function of the Dirac
particles in the scalar and vector central potentials have been
written as $\Psi_{njm\kappa\nu}$ with the following dual symmetry
and the conserved quantum numbers,
\begin{eqnarray}
 \widehat{C} \Psi_{njm\kappa\nu}
 & =  &
 \widetilde{\Psi}_{njm\kappa\nu} = \Psi_{njm\kappa-\nu} ,
 \\
 \widehat{H} \Psi_{njm\kappa\nu}
 & = &
 \nu E_{nj\kappa} \Psi_{njm\kappa\nu} ,
 \\
 \widehat{J}^2 \Psi_{njm\kappa\nu}
 & = &
 j(j+1) \Psi_{njm\kappa\nu} ,
 \\
 \widehat{J}_z \Psi_{njm\kappa\nu}
 & = &
 m \Psi_{njm\kappa\nu} ,
 \\
 \widehat{\kappa}/(j+1/2) \Psi_{njm\kappa\nu}
 & = &
 \kappa \Psi_{njm\kappa\nu} .
\end{eqnarray}

It is remarkable that the anti-symmetry of a system does not lead
to a conserved operator because the anti-symmetry operator is not
commutative but anti-commutative with the Hamiltonian. However, as
the particle-antiparticle space is introduced, the discrete group
$(\mathbb{I},C)$ can be embedded in the enlarged continuous
intrinsic particle-antiparticle space as a discrete subgroup of
the enlarged continuous group with the generators $(\tau_{i})$ and
one can find the invariant operators and the corresponding quantum
numbers. For example, in the cases investigated we do find both
the anti-symmetry operator $C$ and the conserved
particle-antiparticle operator $T_{3}$.


\textit{Summary.} We find that the internal $\gamma $-degrees of
freedom in relativistic quantum mechanics and quantum field theory
is a composite and coherent combination of the spin degrees of
freedom and the particle-antiparticle degrees of freedom. This
coherent combination can only be realized for free particles and
it hides the particle-antiparticle degrees of freedom. As the
coherent composite $\gamma $-degrees of freedom are split by
interactions, the $\gamma $ space is broken into the spin
$\sigma$-space and particle-antiparticle $\tau$-space. In general
cases with interactions, one needs to describe the Dirac particles
in the $\sigma$-space and $\tau$-space separately, which provides
an alternative approach besides the nonlinear Clifford algebra or
Dirac ring. This implies that the splitting of the coherent
composite $\gamma$-space into its constituent $\sigma$-space and
$\tau$-space is just the reflection of the breaking of its
coherence by interactions. The advantage of the splitting is that
it is more ``microscopic'' than Dirac ring and it leads to new and
more detailed information of the wave function and the
corresponding quantum number because it introduces a new intrinsic
space---the particle-antiparticle space. For example, for the
ortho-positronium, quite different from the conventional
description, one can write down explicitly its more detailed wave
function with the quantum numbers in configuration space, spin
space, and particle-antiparticle space as $(l=0,s=1,\tau=0)$, thus
$C$-parity is negative (since the $\tau$ wave function is
anti-symmetric); while for the para-positronium, the quantum
numbers are $(l=0,s=0,\tau=1)$ and $C$-parity is positive (since
the $\tau$ wave function is symmetric). Thus explanation of the
decay properties of positronium is more microscopic and
transparent than the conventional theory~\cite{Lee83}.

\begin{acknowledgments}
S.J.W. and S.G.Z. are indebted to Max-Planck Institute for Nuclear
Physics and particularly to Professor Hans-Christian Pauli for
their hospitality. S.J.W. was partly supported by the National
Natural Science Foundation of China under Grant Nos. 10375039 and
10175029. S.G.Z. was partly supported by the National Natural
Science Foundation of China under Grant No. 10475003, the Major
State Basic Research Development Program of China under contract
No. G20000774 and the Knowledge Innovation Project of Chinese
Academy of Sciences under contract No. KJCX2-SW-N02.

\end{acknowledgments}

\end{document}